\documentclass[12]{amsart}

\usepackage{graphicx}
\usepackage{overpic}
\usepackage{epsfig}

\newcommand{\R}{{\mathbb R}}
\newcommand{\C}{{\mathbb C}}

\renewcommand{\phi}{\varphi}

\newcommand{\al}{\alpha}

\newcommand{\Zn}{Z_{\la_n}}
\newcommand{\Zla}{Z_{\la}}
\newcommand{\Znk}{Z_{\la_{n_k}}}
\newcommand{\Znl}{Z_{\la_{n}^{(\ell)}}}
\newcommand{\lank}{\la_{n_k}}
\newcommand{\Znkl}{Z_{\la_{n_k}^{(\ell)}}}
\newcommand{\lankl}{\la_{n_k}^{(\ell)}}
\newcommand{\lanl}{\la_{n}^{(\ell)}}
\newcommand{\Agmon}{\frac{1}{\pi}|\sqrt{q(z)}| \, |d\gamma|}
\newcommand{\Agmonl}{\frac{1}{\pi}|\sqrt{q(z)}| \, |d\gamma_{\ell}|}
\newcommand{\la}{\lambda}
\newcommand{\ep}{\varepsilon}

\newcommand{\half}{{\frac{1}{2}}}

\newcommand{\haf}{{\frac{1}{2}}}

\renewcommand{\phi}{\varphi}

\newtheorem{theo}{{\sc Theorem}}[section]

\newtheorem{rem}[theo]{{\sc Remark}}

\newtheorem{lem}[theo]{{\sc Lemma}}

\title[Complex zeros of eigenfunctions of 1D Schr\"odinger operators ]{Complex zeros of eigenfunctions of 1D Schr\"odinger operators}

\author{HAMID HEZARI}

\address{Department of Mathematics, Johns Hopkins University, Baltimore, MD
21218, USA}

\email{hhezari@math.jhu.edu}

\date{March, 2007}

\begin{document}

\maketitle

\begin{abstract} In this article we study the semi-classical distribution of the complex zeros of the eigenfunctions
of the 1D Schr\"odinger operators for the class of real polynomial
potentials of even degree, with fixed energy level, $E$. We show
that as $h_n \to 0$ the zeros tend to concentrate on the union of
some level curves $\Re(S(z_m,z))=c_m$ where $S(z_m,z)=\int_{z_m}^z
\sqrt{V(t)-E}\,dt$ is the complex action, and $z_m$ is a turning
point. We also calculate these curves for some symmetric and
non-symmetric one-well and double-well potentials. The example of
the non-symmetric double-well potential shows that we can obtain
different pictures of complex zeros for different subsequences of
$h_n$.


\end{abstract}

\section{Introduction}

This article is concerned with the eigenvalue problem for a
one-dimensional semi-classical Schr\"odinger operator

 \begin{equation} \label{Schro} (- h^2 \frac{d^2}{dx^2}  + V(x)) \psi(x,h) = E(h)
\psi(x,h),\qquad \psi(x,h) \in L^2(\R) \quad h\to 0^+
\end{equation} Using the spectral theory of the Shr\"odinger
operators \cite{BS}, we know that if $\lim_{x\to \underline +
\infty} V(x)=+\infty$ then the spectrum is discrete and can be
arranged in an increasing sequence $E_0(h)< E_1(h)< E_2(h)< \cdots
\uparrow \infty $. Notice that each eigenvalue has multiplicity
one. We let $\{\psi_n(x,h)\}$ be a sequence of eigenfunctions
associated to $E_n(h)$. If we assume the potential $V(x)$ is a
real polynomial of even degree with positive leading coefficient,
then we can arrange the eigenvalues as above and the
eigenfunctions $\psi_n(x,h)$ possess analytic continuations
$\psi_n(z,h)$ to $\C$. Our interest is in the distribution of
complex zeros of $\psi_n(z,h)$ as $h \to 0^+$ when an energy level
$E$ is fixed. The substitutions $\lambda=\frac{1}{h}$, and
$q(x)=V(x)-E$ changes the eigenvalue problem (\ref{Schro}) to the
problem:
\begin{equation} \label{LaSch} y''(x,\la)=\lambda^2 q(x)y(x,\la), \qquad y(x,\la) \in L^2(\R), \qquad \la \to \infty.
\end{equation}
Since $\lim_{x\to \underline + \infty} q(x)=+\infty$, again the
spectrum is discrete and can be arranged as \begin{equation}
\label{SPECTRUM} \la_0<\la_1<\la_2< \cdots <\la_n \cdots \uparrow
\infty.\end{equation} We define the discrete measure $Z_{\la_n}$
by
\begin{equation} \label{ZLa} Z_{\la_n}  =\frac{1}{\la_n}
\sum_{\{z|\; y(z,\la _n)=0 \}} \delta_z. \end{equation}

In this paper we study the limits of weak$^*$ convergent
subsequences of the sequence $\{Z_{\la_n}\}$ as $n \to \infty$. We
say $$ Z_{\la_{n_k}} \longrightarrow Z, \quad (\text{in weak}^* \;
\text{sense})$$ if for every test function $\phi \in
C_c^\infty(\mathbb R^2)$ we have
$$ Z_{\la_{n_k}}(\phi) \longrightarrow Z(\phi).$$ We will call these weak limits, \textit{the zero limit measures}.

Throughout this article we assume that $q(z)$ has simple zeros. We
may be able to extend the results in the case of multiple turning
points using the methods in \cite{F1,O2} on the asymptotic
expansions around multiple turning points. Notice that $q(x)$ has
to change its sign on the real axis, because if it is positive
everywhere then (\ref{LaSch}) does not have any solution in
$L^2(\R)$. Hence $q(z)$ has at least two simple real zeros. We say
$q(x)$ is a one-well potential if it has exactly two (simple) real
zeros and a double-well potential if it has exactly four (simple)
real zeros. One of our results is

\begin{theo} \label{FormOfZeros} Let $q(z)$ be a real polynomial of even degree with
positive leading coefficient. Then every weak limit $Z$ (zero
limit measure) of the sequence $\{Z_{\la_n}\}$ is of the form

\begin{equation} \label{Z} Z= \Agmon, \end{equation}
$$|d\gamma|=|\gamma'(t)| \, dt, $$
where $\gamma$ is a union of finitely many smooth connected curves
$\gamma_m$ in the plane. For each $\gamma_m$ there exists a
constant $c_m$, a canonical domain $D_m$ and a turning point $z_m$
on the boundary of $D_m$ such that $\gamma_m$ is given by

\begin{equation} \Re(S(z_m,z))=c_m, \qquad z \in D_m,
\end{equation} where $S(z_m,z)=\int_{z_m}^z \sqrt{q(t)} \, dt$ and
the integral is taken along any path in $D_m$ joining $z_m$ to
$z$.(See section 2, for the definitions)

\end{theo}

This theorem shows that if $Z$ in (\ref{Z}) is the limit of a
subsequence $\{Z_{\la_{n_k}}\}$, then the complex zeros of
$\{y(z,\la_{n_k})\}$ tend to concentrate on $\gamma$ as $k \to
\infty$ and in the limit they cover $\gamma$. The factor
$|\sqrt{q(z)}|=|\sqrt{V(z)-E}|$ indicates that the limit
distribution of the zeros on $\gamma$ is measured by the Agmon
metric. We call the curves $\gamma_m$ the \textit{zeros lines} of
the limit $Z$. The next question after seeing Theorem
\ref{FormOfZeros} is "what are all the possible zero limit
measures and corresponding zero lines for a given polynomial
$q(z)$?" We answer this question for some one-well and double-well
potentials. But before stating these results let us mention some
background and motivation for the problem.

Form the classical Sturm-Liouville theory we know everything about
the real zeros of solutions of (\ref{LaSch}). We know that on a
classical interval (i.e. an interval where $q(x)<0$), every
real-valued solution $y(x, \la)$ of (\ref{LaSch}) (not necessarily
$L^2$-solution) is oscillatory and becomes highly oscillatory as
$\la \to \infty$. In fact the spacing between the real zeros on a
classical interval is $\frac{\pi}{\la}$. On the other hand there
is at most one real zero on each connected forbidden interval
where $q(x)>0$. This shows that every limit $Z$ in (\ref{Z}) has
the union of classical intervals in its support.

It turns out that other than the harmonic oscillator $q(z)=z^2
-a^2$ where the eigenfunctions do not have any non-real zeros, the
complex zeros are more complicated. It is easy to see that when
$q(z)=z^4+az^2+b$, the eigenfunctions have infinitely many zeros
on the imaginary axis. For $q(z)=z^4 \underline + a^4$, Titchmarsh
in \cite{T} made a conjecture that all the non-real zeros are on
the imaginary axis. This conjecture was proved by Hille in
\cite{H1}. In general one can only hope to study the asymptotics
of large zeros of $y(z,\la)$ rather than finding the exact
locations of zeros. The asymptotics of zeros of solutions to
(\ref{LaSch}) for a fixed $\la$ and large $z$, have been
extensively studied mainly by E. Hille, R. Nevanlinna, H. Wittich
and S. Bank (see \cite{N}, \cite{W}, \cite{B}). But it seems the
semi-classical limit of complex zeros has not been studied in the
literature, at least not from the perspective that was mentioned
in Theorem \ref{FormOfZeros}, which is closely related to the
quantum limits of eigenfunctions. This problem was raised around
fifteen years ago when physicists were trying to find a connection
between eigenfunctions of quantum systems and the dynamics of the
classical system. It was noticed that for the ergodic case the
complex zeros tend to distribute uniformly in the phase space but
for the integrable systems the zeros tend to concentrate on
one-dimensional lines. An article which made this point and
contains very interesting graphics is \cite{LV}. The problem of
complex zeros of complexified eigenfunctions and relations to
quantum limits is suggested by S. Zelditch mainly in \cite{Z}.
There, the author proves that if a sequence $\{ \phi_{\la_n} \}$
of eigenfunctions of the Laplace-Beltrami operator on a real
analytic manifold $M$ is quantum ergodic then the sequence
$\{Z_{\la_n}\}$ of zero distributions associated to the
complexified eigenfunctions $\{\phi_{\la_n}^{\mathbb C}\}$ on
$M^{\mathbb C}$, the complexification of $M$, is weakly convergent
to an explicitly calculable measure. A natural problem is to
generalize the results in \cite{Z} for Schr\"odinger
eigenfunctions on real analytic manifolds. This is indeed a
difficult problem. Perhaps the first step to study such a problem
is to consider the one dimensional case which we do in this paper.
The main reason to study the complex zeros rather than the real
zeros is that the problem is much easier in this case (in higher
dimensions). For example in studying the zeros of polynomials as a
model for eigenfunctions, the Fundamental Theorem of Algebra and
Hilbert's Nullstellensatz are two good examples of how the complex
zeros are easier and somehow richer. See \cite{Z1} for some
background of the problem and some motivation in higher
dimensions.

One of our results is that for a symmetric quartic oscillator the
full sequence $\{Z_{\la_n}\}$ is convergent, i.e. there is a
unique zero limit measure. Here, by $q(z)$ being symmetric we mean
that after a translation on the real axis, q(z) is an even
function. But for a non-symmetric quartic oscillator there are at
least two zero limit measures. The Stokes lines play an important
role in the description of the zero lines. In fact the infinite
zero lines are asymptotic to Stokes lines. This fact was observed
in \cite{B}.

Our proofs are elementary. We use the complex WKB method,
connection formulas and asymptotics of the eigenvalues by
Fed\"oryuk in \cite{F1}.

At the end of the paper (\ref{Remark}) we will briefly mention
some interesting examples of one-well and double-well potentials
where deg$(q(z))=4,6$.





\begin{theo} \label{2well}
Let $q(z)=(z^2-a^2)(z^2-b^2)$, where $0<a<b$. Then as $n \to
\infty$
$$ Z_{\la_n} \longrightarrow
\Agmon, \qquad \gamma= (a,b) \cup (-b,-a) \cup (-\infty i,+\infty
i).$$

\end{theo}

Notice in Theorem \ref{2well} we can express $\gamma$ by three
equations
$$\gamma=\{\Re S(a,z)=0\}\cup \{\Re S(-a,z)=0\} \cup \{\Re
S(a,z)=-\half \xi\},$$ where $\xi=\int_{-a}^{a} \sqrt{q(t)}\,dt$,
and each equation is written in some canonical domain.

\begin{theo} \label{1well}
Let $q(z)=(z^2-a^2)(z^2+b^2)$, where $a,b>0$. Then as $n \to
\infty$ $$ Z_{\la_n} \longrightarrow \Agmon, \qquad \gamma= (-a,a)
\cup (bi,+\infty i) \cup (-\infty i, -bi).$$

We also note that in Theorem \ref{1well} we can express $\gamma$
by three equations
$$\gamma=\{\Re S(a,z)=0\}\cup \{\Re S(bi,z)=0\} \cup \{\Re
S(-bi,z)=0\},$$ where each equation is written in some canonical
domain.

\end{theo}

Theorems \ref{2well} and \ref{1well} state that for a symmetric
quartic polynomial there is a unique zero limit measure. This is
not always the case when $q(z)$ is not symmetric. Let
$q(z)=(z-a_0)(z-a_1)(z-a_2)(z-a_3)$ where $a_0<a_1<a_2<a_3$. Using
the quantization formulas (for example in \cite{S,F1}), we have
two sequences of eigenvalues
\begin{equation} \label{Eigen1} \la_{n}^{(1)}=\frac{2n+1}{2\al_1} \pi +
O(\frac{1}{n}), \qquad \al_1=\int_{a_0}^{a_1}
|\sqrt{q(t)}|dt,\end{equation}
\begin{equation} \label{Eigen2} \la_{n}^{(2)}=\frac{2n+1}{2\al_2} \pi +
O(\frac{1}{n}), \qquad \al_2=\int_{a_2}^{a_3} |\sqrt{q(t)}|dt.
\end{equation}

Now with this notation we have the following theorem:

\begin{theo} \label{Nonsymm} Let $q(z)=(z-a_0)(z-a_1)(z-a_2)(z-a_3)$, where
$a_0<a_1<a_2<a_3$ are real numbers. Then
\begin{itemize}

\item[1.] if $\frac{\al_1}{\al_2}$ is irrational then for each
$\ell \in \{1,2\}$ there is a full density subsequence
$\{\lankl\}$ of $\{\lanl\}$ such that $$ \Znkl \longrightarrow
\Agmonl \; , $$

where

\begin{equation}\label{gammal} \begin{array}{ll}
\gamma_1=(a_0,a_1) \cup (a_2,a_3) \cup \{\Re(S(a_2,z)=0\},
\\ \\ \gamma_2=(a_0,a_1) \cup (a_2,a_3) \cup
\{\Re(S(a_1,z)=0\}. \end{array} \end{equation}

\item[2.] if $\frac{\al_1}{\al_2}$ is rational and of the form
$\frac{2r_1}{2r_2+1}$ or $\frac{2r_1+1}{2r_2}$  then for each
$\ell \in \{1,2\}$ $$ \Znl \longrightarrow \Agmonl.$$

\item[3.] if $\frac{\al_1}{\al_2}$ is rational and of the form
$\frac{2r_1+1}{2r_2+1}$ where $\gcd(2r_1+1,2r_2+1)=1$, then for
each $\;\ell \in \{1,2\}$ there exists a subsequence $\{\lankl\}$
of $\{\lanl\}$ of density $\frac{2r_{\ell}}{2r_{\ell}+1}$ such
that
$$ \Znkl \longrightarrow \Agmonl.$$ In fact $\{\lankl\}=\{\lanl |
2n+1\neq 0 \;(\text{mod}\; 2r_{\ell}+1)\}$.
\end{itemize}
\end{theo}

\begin{figure}
\begin{overpic}[scale=0.7]{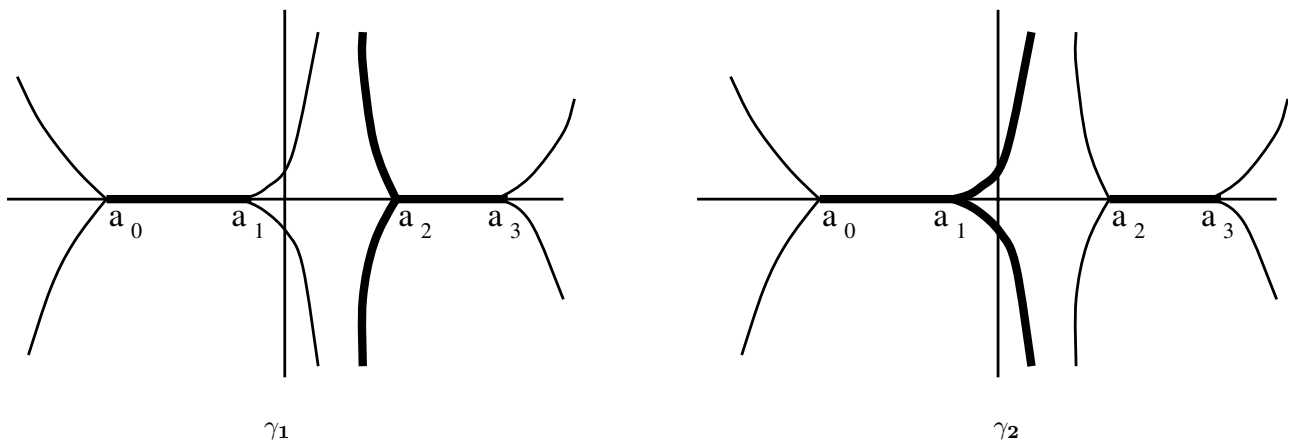}
\put(20,8){$\mathbf{{\gamma}_1}$}
\put(77,8){$\mathbf{{\gamma}_2}$}
\end{overpic}
\caption{\label{GammaEll} The zero lines of the non-symmetric
quartic oscillator. The thick lines are the zeros lines $\gamma_1$
and $\gamma_2$ given by (\ref{gammal}).}

\end{figure}

Figure (\ref{GammaEll}) shows $\gamma_1$ and $\gamma_2$ defined in
(\ref{gammal}). As we see the zero lines here are some of the
Stokes lines. We should mention that in Theorem \ref{Nonsymm},
when $\frac{\al_1}{\al_2}$ is irrational or of the form
$\frac{2r_1+1}{2r_2+1} \neq 1$, we do not know what happens to the
rest of the subsequences. There might be some exceptional
subsequences (of positive density in the case
$\frac{\al_1}{\al_2}=\frac{2r_1+1}{2r_2+1}$) for which the zero
lines are different from $\gamma_{\ell}$. As we saw in Theorem
\ref{2well}, this is the case for the symmetric double-well
potential when $\frac{\al_1}{\al_2}=1$. We probably need a more
detailed analysis of the eigenvalues in order to answer this
question.

\begin{rem}

In this remark we mention some very recent results of A. Eremenko,
A. Gabrielov, B. Shapiro in \cite{EGS1} and \cite{EGS2}, and
compare them to ours as the interests and the approaches in these
articles are very similar to ours.

1. Theorems \ref{2well} and \ref{1well} do not say anything about
the exact location of the zeros but they only state that as $n \to
\infty$ the zeros approach to $\gamma$ with the distribution law
in [\ref{Z}]. It is easy to see that for both of these symmetric
cases, for each $n$, all the zeros of $y(z,\la_n)$ except finitely
many of them are on $\gamma$. In \cite{EGS1}, the authors prove
that for the solutions of the equation \begin{equation}
\label{EGSequation} -y''+P(x)y=\la y, \qquad y\in L^2(\mathbb R),
\end{equation}

where $P(x)$ is an even real monic polynomial of degree 4, all the
zeros of $y(z)$ belong to the union of the real and imaginary
axis. This result indeed implies that for all $n$, all the zeros
of $y(z,\la_n)$ in Theorem \ref{2well} and Theorem \ref{1well} are
on the corresponding $\gamma$.
\\

2. In \cite{EGS2}, the authors show that the complex zeros of the
scaled eigenfunctions $Y_n(z)=y(\la_n^{1/d}z)$ of
$(\ref{EGSequation})$, where $d=$degree$(P(x))$, have a unique
limit distribution in the complex plane as $\la_n \to \infty$. The
scaled eigenfunctions satisfy an equation of the form
$$Y_n''(z)=k_n^2(z^d-1+o(1))Y_n(z), \qquad k_n \to \infty.$$

The main reason that they could establish a uniqueness result for
the limit distribution of complex zeros of $Y_n(z)$ is due the
special structure of the Stokes graph of the polynomial $z^d-1$
which is proved in Theorem 1 in \cite{EGS2}.

\end{rem}

\section{A Review of the Complex WKB Method}

To prove the theorems we first review some basic definitions and
facts about complex WKB method. We follow \cite{F1}. See
\cite{O1,S,EF} for more references on this subject.
\\

We consider the equation \begin{equation} \label{complexified}
y''(z,\la)=\lambda^2 q(z)y(z,\la), \qquad \qquad \quad \lambda \to
\infty,
\end{equation} on the complex plane $\C$, where $q(z)$ is a
polynomial with simple zeros.

\subsection{Stokes lines and Stokes graphs}A zero $z_0$, of $q(z)$ is called a turning point. We let
$S(z_0,z)=\int_{z_0}^z \sqrt{q(t)}dt$. This function, is in
general, a multi-valued function. The maximal connected component
of the level curve $\Re(S(z_0,z))=0$ with initial point $z_0$ and
having no other turning points are called the Stokes lines
starting from $z_0$. Stokes lines are independent of the choice of
the branches for $S(z_0,z)$. The union of the Stokes lines of all
the turning points is called the Stokes graph of
(\ref{complexified}).

\begin{figure}
\begin{center} \includegraphics[scale=0.5]{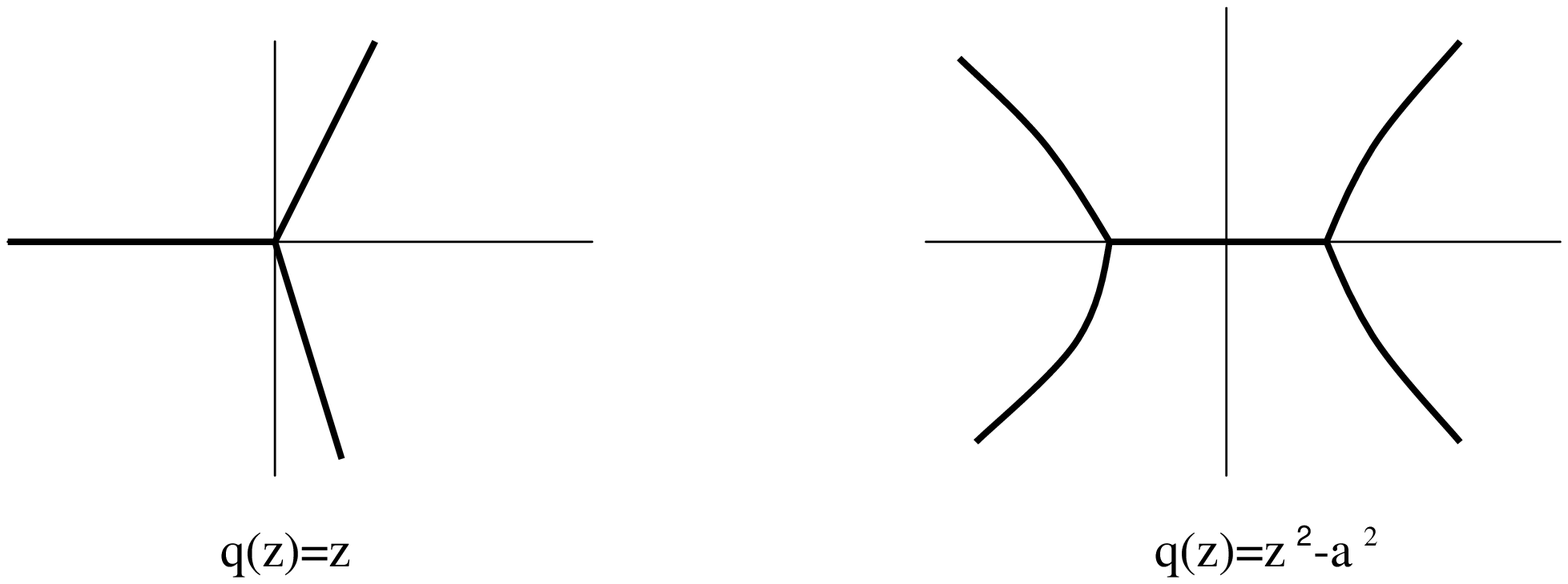}\end{center}
\begin{center} \includegraphics[scale=0.5]{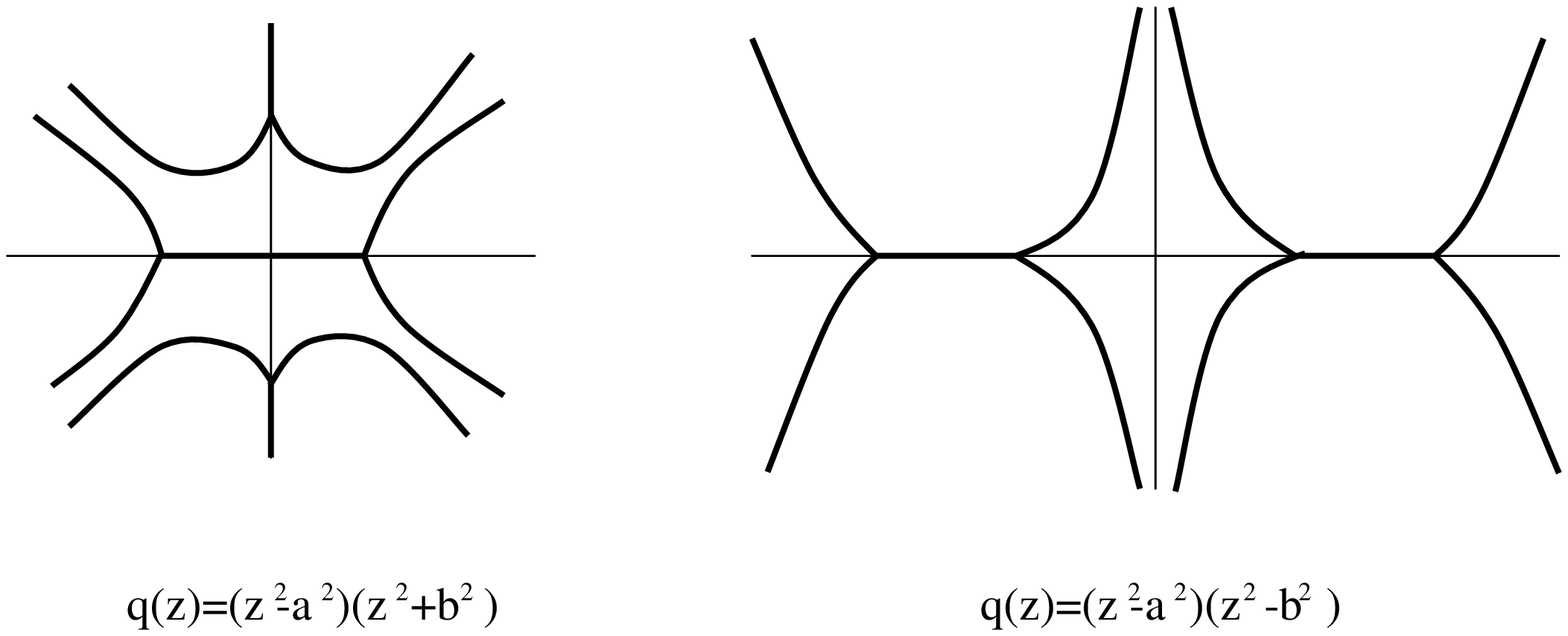}\end{center}
\begin{center} \includegraphics[scale=0.5]{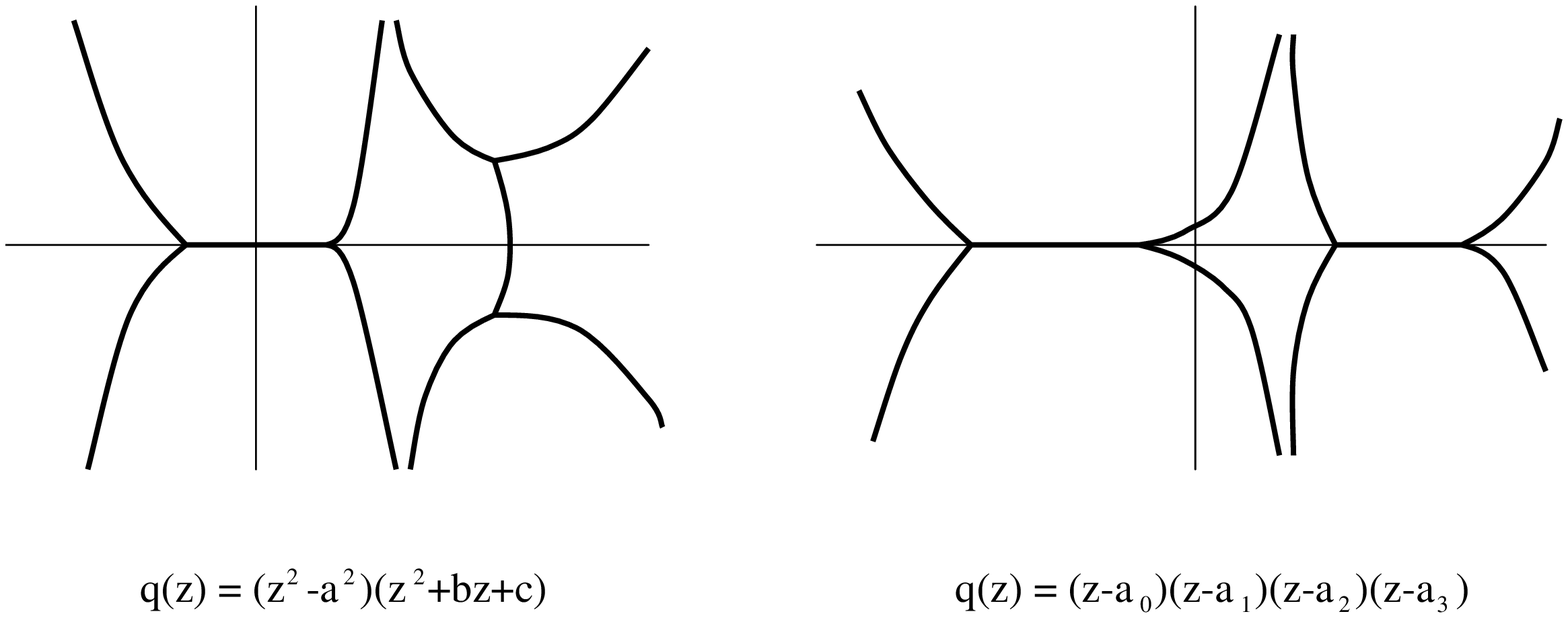}\end{center}
\caption{\label{StokesGraph} Stokes lines for some polynomials}
\end{figure}

Figure (\ref{StokesGraph}) shows the Stokes graphs of many
polynomials.

Since the turning points are simple, from each turning point three
Stokes lines emanate with equal angles. In general if $z_0$ is a
turning point of order $n$, then $n+2$ Stokes lines with equal
angles emanate from $z_0$.

\subsection{Canonical Domains, Asymptotic Expansions}

Since $q(z)$ is a polynomial, the Stokes graph divides the complex
plane into two type of domains:

\begin{itemize}
\item[1.]Half-plane type: A simply connected domain $D$ which is bounded by
Stokes lines is a half-plane type domain if under the map $S=S(z_0,z)$, it is
biholomorphic to a half-plane of the form $\Re S>a$ or $\Re S<a$. Here $z_0$ is
a turning point on the boundary of $D$.

\item[2.] Band-type: $D$ as above, is of band-type if under $S$, it is
biholomorphic to a band of the form $a<\Re S<b$.

\end{itemize}

A domain $D$ in the complex plane is called canonical if
$S(z_0,z)$ is a one-to-one map of D onto the whole complex plane
with finitely many vertical cuts such that none cross the real
axis. A canonical domain is the union of two half-plane type
domains and some band type domains. For example, the union of two
half-plane type domains sharing a Stokes line is a canonical
domain.

Let $\varepsilon >0$ be arbitrary. We denote $D^{\varepsilon}$ for
the pre-image of $S(D)$ with $\varepsilon$-neighborhoods of the
cuts and $\varepsilon$-neighborhoods of the turning points
removed. A canonical path in $D$ is a path such that $\Re(S)$ is
monotone along the path. For example, the anti-Stokes lines (lines
where, $\Im(S)=0$), are canonical paths. For every point $z$ in
$D^{\varepsilon}$, there are always canonical paths
$\gamma^{-}(z)$ and $\gamma^{+}(z)$  from $z$ to $\infty$, such
that $\Re S \downarrow -\infty$ and $\Re S \uparrow \infty$,
respectively.
\\
Now we have the following fact:

With $D$, $\gamma ^+$, $\gamma^-$ as above, to within a multiple
of a constant, equation (\ref{complexified}) has a unique solution
$y_1(z,\lambda)$, such that \begin{equation} \label{y1} \lim_{z
\to \infty, z \in \gamma^{-}} y_1(z,\lambda)=0, \qquad \quad \Re
S(z_0,z) \downarrow -\infty,
\end{equation} and a unique solution $y_2(z,\lambda)$ (up to a constant
multiple), such that
\begin{equation} \label{y2} \lim_{z \to \infty, z \in \gamma {+}} y_2(z,\lambda)=0, \qquad \quad \Re S(z_0,z) \uparrow
\infty. \end{equation} The solutions $y_1$ and $y_2$ in (\ref{y1})
and (\ref{y2}) have uniform asymptotic expansions in
$D^{\varepsilon}$ in powers of $\frac{1}{\la}$. Here we only state
the principle terms:

\begin{equation} \label{y1princ} y_1(z,\lambda)=q^{\frac{-1}{4}}(z)e^{\lambda S(z_0,z)}(1+\ep_1(z,\la)) \qquad \lambda \to \infty,\end{equation}
\begin{equation} \label{y2princ} y_2(z,\lambda)= q^{\frac{-1}{4}}(z)e^{-\lambda S(z_0,z)}(1+\ep_2(z,\la)) \qquad \lambda \to \infty,\end{equation}
where
\begin{equation} \label{ep1} \ep_1(z,\la)=O(\frac{1}{\la}), \qquad
\text{uniformly in} \; D^{\ep},\quad \la \to \infty,\end{equation}
\begin{equation} \label{ep2} \ep_2(z,\la)=O(\frac{1}{\la}), \qquad
\text{uniformly in} \; D^{\ep},\quad \la \to \infty.\end{equation}

Notice that the equalities (\ref{ep1}) and (\ref{ep2}) would not
necessarily be uniformly in $D^{\ep}$ if $q(z)$ was not a
polynomial.

\subsection {Elementary Basis} Let $D$ be a canonical
domain, $l$ a Stokes line in $D$, and $z_0 \in l$ a turning point.
We use the triple $(D,l,z_0)$ to denote this data. We select the
branch of $S(z_o,z)$ in $D$ such that $\Im S(z_0,z)>0$ for $z \in
l$. The elementary basis $\{ u(z),v(z) \}$ associated to
$(D,l,z_0)$ is uniquely defined by \begin{equation} \label{basis}
\left\{\begin{array}{ll} u(z,\lambda)=cy_1(z,\la), \quad
\text{and} \quad v(z,\lambda)=cy_2(z,\la), \\ \\ |c|=1, \;
\text{and} \; \arg (c)=\lim_{z\to z_0, z \in l} \arg (q^{1/4}(z)),
\end{array} \right.
\end{equation}

where $y_1(z,\la)$, $y_2(z,\la)$ are given by (\ref{y1princ}) and
(\ref{y2princ}).

\subsection{Transition Matrices} Assume $(D,l,z_0)_j$ and $(D,l,z_0)_k$ are two
triples and $\beta _j=\{ u_j,v_j \}$ and $\beta _k=\{ u_k,v_k \}$ their
corresponding elementary basis. The matrix $\Omega _{jk}(\la)$ which changes
the basis $\beta _j$ to $\beta _k$ is called the transition matrix from $\beta
_j$ to $\beta _k$.

Fed\"oryuk, in [EF], introduced three types of transition matrices
that he called elementary transition matrices, and he proved that
any transition matrix is a product of a finitely many of these
elementary matrices. The three types are

\begin{itemize}
\item [1)] $(D,l,z_1) \mapsto (D,l,z_2)$. This is the transition
from one turning point to another along a finite Stokes line
remaining in the same canonical domain $D$. The transition matrix
is given by \begin{equation} \label{Transition1}
\Omega(\la) = e^{i \phi} \left(%
\begin{array}{cc}
  0 & e^{-i\lambda \alpha} \\
  e^{i\lambda \alpha} & 0 \\
\end{array}%
\right), \qquad \alpha = |S(z_1,z_2)|, \quad e^{i \phi
}=\frac{c_2}{c_1}. \end{equation}

\item [2)] $(D,l_1,z_1) \mapsto (D,l_2,z_2)$. Here the rays
$S(l_1)$ and $S(l_2)$ are directed to one side. This is the
transition from one turning point to another along an anti-Stokes
line, remaining in the same domain $D$. The transition matrix is
\begin{equation} \label{Transition2}
\Omega(\la) = e^{i \phi} \left(%
\begin{array}{cc}
  e^{-\lambda a} & O \\
  0 & e^{\lambda a} \\
\end{array}%
\right), \qquad a=|S(z_1,z_2)|, \quad e^{i \phi }=\frac{c_2}{c_1}.
\end{equation}

\item[3)]$(D_1,l_1,z_0) \mapsto (D_2,l_2,z_0)$ This is a simple
rotation around a turning point $z_0$ so that $D_1$ and $D_2$ have
a common sub-domain. More precisely, let $\{l_j; j=1,2,3\}$ be the
Stokes lines starting at $z_0$ and ordered counter-clockwise so
that $l_{j+1}$ is located on the left side of $l_j$. We choose the
canonical domain $D_j$ so that the part of $D_j$ on the left of
$l_j$ equals the part of $D_{j+1}$ on the right of $l_{j+1}$.
Then \begin{equation} \label{AlphaJ} \left\{ \begin{array}{ll} \Omega _{j,j+1}(\la) =e^{- \frac{\pi }{6}} \left(%
\begin{array}{cc}
  0 & \alpha _{j,j+1}^{-1}(\la) \\
  1 & i\alpha _{j+1,j+2}(\la) \\
\end{array}\right),%
 \\ \\ \alpha _{j,j+1}(\lambda )= 1+O(\frac{1}{\la}), \quad 1\leq j
 \leq 3,
\\ \\ \al_{1,2}(\la) \al_{2,3}(\la) \al_{3,1}(\la)=1, \quad
\text{and} \quad \alpha _{j,j+1}(\la)\alpha _{j+1,j}(\la)=1.
\end{array} \right.  \end{equation}

\end{itemize}

\subsection{Polynomials with real coefficients} We finish this section with a review of some properties of the Stokes lines and
transition matrices in (\ref{AlphaJ}) when the polynomial $q(z)$
has real coefficients.

\begin{itemize}

\item [1)] The turning points and Stokes lines are symmetric about
the real axis. If $x_1<x_2$ are two real turning points and
$q(x)<0$ on the line segment $l=[x_1,x_2]$, then $l$ is a Stokes
line (See Figure \ref{StokesGraph}). Similarly, if $q(x)>0$ on
$l$, then $l$ is an anti-Stokes line. Let $x_0$ be a simple
turning point on the real axis, and let $l_0, l_1, l_2$ be the
Stokes lines starting at $x_0$. Then one of the Stokes lines, say
$l_0$, is an interval of the real axis, and $l_2= \bar {l_1}$. The
Stokes lines $l_1$ and $l_2$ do not intersect the real axis other
than at the point $x_0$. If a Stokes line $l$ intersects the real
axis at a non-turning point, then $l$ is a finite Stokes line and
it is symmetric about the real axis.

If $\lim _{x \to \infty} q(x)=\infty$, and $x^{+}$ is the the
largest zero of $q(x)$, and ${l_0,l_1,l_2}$ are the corresponding
Stokes lines, then there is a half type domain $D^{+}$ such that
$$ [x^{+},+\infty] \subset D^{+}, \qquad D^{+}=\overline{D^{+}}, \quad l_1 \cup l_2 \subset
\partial D^{+}.$$ Clearly $[x_0,+\infty ]$ is an anti-Stokes line and
$S(x_0,\infty)=\infty$. By (\ref{y1}), there exists a unique
solution $y^{+}(z,\lambda)$ such that $$ \lim _{x \to
\infty}y^{+}(x,\lambda)=0.$$ Similarly by (\ref{y2}) if $\lim _{x
\to -\infty} q(x)=\infty$ and $x^{-}$ is the smallest root of
$q(x)$, and $D^{-}$ a half type domain containing $[-\infty,
x_0]$, there exists a unique solution $y^{-}(z, \la)$ such that $$
\lim _{x \to -\infty}y^{-}(x,\lambda)=0.$$

Therefore if $y(x,\lambda)$ is an $L^2$-solution to (\ref{LaSch})
, then for some constants $c^{+}$, $c^{-}$
$$y(x,\la)=c^{+}y^{+}(x,\lambda)=c^{-}y^{-}(x,\lambda).$$
Now let $\Omega _{+,-}(\la)$ be the transition matrix connecting
$D^{+}$ to $D^{-}$ and let

$$ \left(%
\begin{array}{c}
  a(\la) \\
  b(\la)\\
\end{array}%
\right)=\Omega _{+,-}(\la)\left(%
\begin{array}{c}
  0 \\
  1 \\
\end{array}%
\right) .$$

The fact that $y^{+}(x,\lambda)$ is a constant multiple of
$y^{-}(x,\lambda)$ is equivalent to \begin{equation} \label{bla}
b(\la)=0,
\end{equation}

which is the equation that determines the eigenvalues $\la_n$. To
calculate $\Omega _{+,-}(\la)$ and hence $b(\la)$ we have to write
this matrix as a product of finitely many elementary transition
matrices connecting $D^{+}$ to $D^{-}$.

\item[2)]When the polynomial $q(z)$ has real coefficients, the
transitions matrices in (\ref{AlphaJ}) have some symmetries. Let
$x_0$ be a simple turning point and $q(x)>0$ on the interval
$[x_0,b]$. We index the Stokes lines $l_0, l_1,l_2$ as in Figure
(\ref{Alpha12}). We define the canonical domains $D_0, D_1, D_2$
by their internal Stokes lines and their boundary Stokes lines as
the following
$$ D_0=\overline{D_0}, \qquad l_0 \subset D_0, l_1 \cup l_2 \subset \partial D,$$
$$ [x_0,b] \subset D_1, \qquad l_0 \cup l_2 \subset \partial D_1,$$
$$ D_2=\overline D_1.$$

\begin{figure}
\begin{center}\includegraphics[scale=0.5]{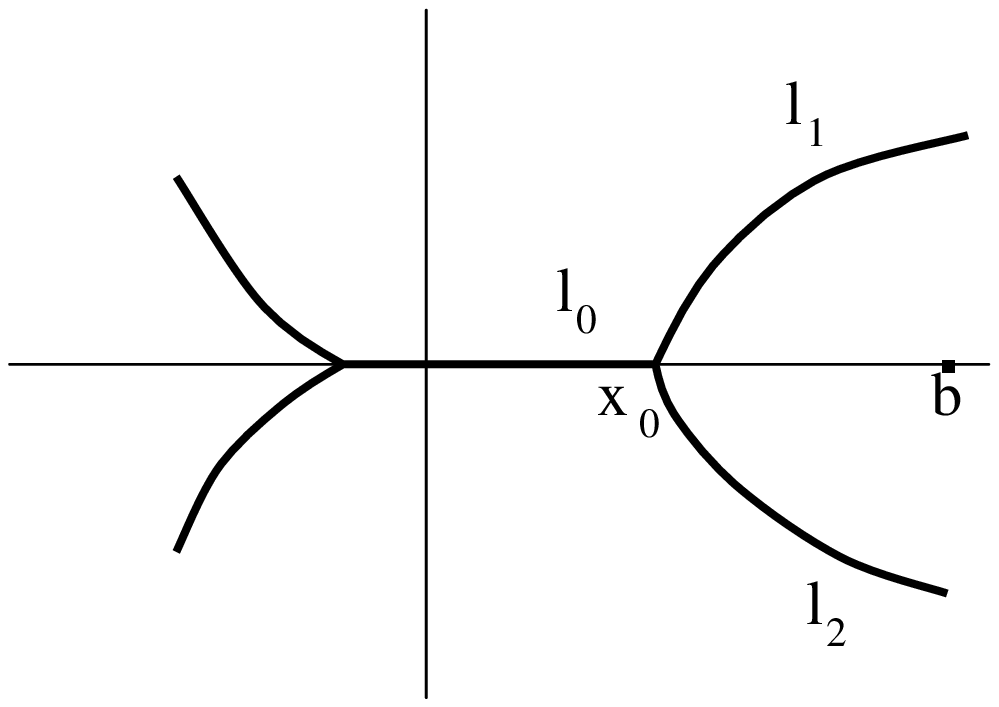}\end{center}
\caption{\label{Alpha12}}
\end{figure}

Now with the same notation as in (\ref{AlphaJ}), we have
\begin{equation} \label{Alpha12} \alpha _{0,1}= \overline {\alpha_ {0,2}}, \qquad
|\alpha _{1,2}|=1.
\end{equation}

\end{itemize}

\section{Proofs of the Theorems}

First of all we have the following lemma:

\begin{lem} \label{lem} Let $T=(D,l,z_0)$ be a triple as in section 2.3 and let
$\{u(z,\la),v(z,\la)\}$ be the elementary basis associated to $T$
in (\ref{basis}). We write $y(z,\la_n)$ in this basis as
\begin{equation} \label{alabla}
y(z,\la_n)=a(\la_n)u(z,\la_n)+b(\la_n)v(z,\la_n). \end{equation}

If $\{\lank\}$ is a subsequence of $\{\la_n\}$ such that the limit
\begin{equation} \label{t} t=\lim_{k\to \infty} \frac{1}{2\lank} \log
|\frac{b(\lank)}{a(\lank)}|, \end{equation} exists, then in
$D^{\ep}$ we have $$\Znk \longrightarrow \Agmon, \qquad
\gamma=\{z\in D|\;\Re S(z_0,z)=t\}.$$ The last expression means
that for every $\phi \in C_c^\infty(D^{\ep})$ we have $$ \Znk
(\phi) \to \frac{1}{\pi}\int_{\gamma}
\phi(z)|\sqrt{q(z)}||d\gamma|.$$
\end{lem}

\textsc{Proof of Lemma}. For simplicity we omit the subscript
$n_k$ in $\lank$, but we remember that the limit in (\ref{t}) is
taken along $\lank$. Using (\ref{alabla}), (\ref{basis}),
(\ref{y1princ}), and (\ref{y2princ}), the equation $y(z,\la)=0$ in
$D^{\ep}$, is equivalent to
\begin{equation} \label{newequation} S(z_0,z)-\frac{1}{2\la} \log
(\frac{1+\ep_1(z,\la)}{1+\ep_2(z,\la)})=\frac{1}{2\la}\log \mid
\frac{b(\la)}{a(\la)} \mid + i(\frac{2k+1}{2\la}\pi+
\frac{1}{2\la}\arg(\frac{b(\la)}{a(\la)})), \qquad k \in \mathbb
Z.\end{equation}

where we have chosen $\log z=\log r+i\theta$, $-\pi < \theta
<\pi.$ We use $\tilde{S}(z)$ for the function on the left hand
side of (\ref{newequation}) and $a_k$ for the sequence of complex
numbers on the right hand side. As we see $\tilde{S}(z)$ is the
sum of the biholomorphic function $S(z_0,z)$ and the function

\begin{equation} \label{smallterm} \mu (z,\la):=-\frac{1}{2\la} \log
(\frac{1+\ep_1(z,\la)}{1+\ep_2(z,\la)})=O(\frac{1}{\la ^2}),
\qquad  \text{uniformly in} \; D^{\ep},\; \text{by} \;
(\ref{ep1}), (\ref{ep2}). \end{equation}

Now suppose $\phi \in C_c^\infty(D^{\ep})$ and
$K=\text{supp}(\phi)$. We also define $K'=S(K)$ where
$S(z)=S(z_0,z)$. Without loss of generality we can assume that
$\{x=t\}\cap \text{int}(K')$ is a connected subset of the vertical
line $x=t$, because we can follow the same argument for each
connected component. Now let $s=\text{length}(\{x=t\}\cap
\text{int}(K'))$. It is clear that because of (\ref{t})
\begin{equation}\label{N} N:=\#\{\; a_k \in \text{int}(K')\} \sim
s\la.\end{equation}

We call this finite set $\{a_k\}_{m+1\leq k \leq m+N}$. Now let $K
\subset V \subset D^{\ep}$ be an open set with compact closure in
$D^{\ep}$. We choose $\la$ large enough such that $$ |\mu(z,\la)|
< |S(z_0,z)-a| \qquad \forall \; a \in K', \quad \forall \; z\in
\partial V.$$

Since $S$ is a biholomorphic map, by Rouch\' e's theorem the
equation $$ \tilde{S}(z)=a_k, \qquad m+1 \leq k \leq m+N , $$ has
a unique solution $z_k$ in $V$ for each $k$. Now by (\ref{t}),
(\ref{newequation}), and (\ref{smallterm}), we have
$$ z_k=S^{-1}\big(\frac{1}{2\la}\log \mid
\frac{b(\la)}{a(\la)} \mid -\mu(z_k,\la))+ i(\frac{2k+1}{2\la}\pi+
O(\frac{1}{\la}))\big) $$

$$=S^{-1}\big(t+o(1)+
i(\frac{2k+1}{2\la}\pi-\Im(\mu(z_k,\la))+O(\frac{1}{\la}))\big).$$

It follows that $$\Zla (\phi)=\frac{1}{\la} \sum_{k=m+1}^{m+N}
\phi (z_k)=\frac{1}{\la} \sum_{k=m+1}^{m+N} (\phi \circ
S^{-1})\big(t+o(1)+
i(\frac{2k+1}{2\la}\pi-\Im(\mu(z_k,\la))+O(\frac{1}{\la}))\big).$$

Using the mean value theorem on the $x-$axis and (\ref{N}), we
obtain
\begin{equation} \label{Riemann} \lim_{\la \to \infty} \Zla
(\phi) =\lim_{\la \to \infty}\frac{1}{\la} \sum_{k=m+1}^{m+N}
\{[(\phi \circ S^{-1})\big(t+
i\big(\frac{2k+1}{2\la}\pi-\Im(\mu(z_k,\la))+
O(\frac{1}{\la})\big)\big)]+o(1)\}.\end{equation}

Because of (\ref{smallterm}), we know that
$\Im(\mu(z_k,\la))=O(\frac{1}{\la^2})$ uniformly in $k$. Therefore
the set
$$\wp=\{(t,\frac{2k+1}{2\la}\pi-\Im(\mu(z_k,\la))+
O(\frac{1}{\la}))| \; m+1 \leq k \leq m+N\}$$ is a partition of
the vertical interval $\{x=t\}\cap \text{int}(K')$ with
$\text{mesh}(\wp) \to 0$ as $\la \to \infty$. This together with
(\ref{Riemann}) implies that

$$ \lim_{\la \to \infty} \Zla (\phi) = \frac{1}{\pi} \int_{\{x=t\}}
\phi \circ S^{-1} dy.$$

Now, if in the last integral we apply the change of variable $z
\mapsto S(z)$, then by the Cauchy-Riemann equations for $S$, we
obtain
$$\frac{1}{\pi} \int_{\{x=t\}} \phi \circ S^{-1} dy= \frac{1}{\pi}
\int_{\{\Re S(z)=t\}} \phi (z) |\sqrt{q(z)}| \, |d\gamma|.$$

This proves the Lemma.
\\

\textsc{Proof of Theorem \ref{FormOfZeros}}. First of all, we
cover the plane by finitely many canonical domains $D_m$. Let $\ep
>0$ be sufficiently small as before.  Assume $\{\Znk \}$ is a
weak$^*$ convergent subsequence converging to a measure $Z$.
Clearly $\{\Znk \}$ converges to $Z$ in each $D_m^{\ep}$. We claim
that the limit (\ref{t}) exists for every triple
$T_m=(D_m,z_m,l_m)$. This is clear from Lemma \ref{lem}. This is
because if in (\ref{t}) we get two distinct limits $t_1$ and $t_2$
for two subsequences of $\{\Znk \}$, then we get two corresponding
distinct limits $Z_1$ and $Z_2$ which contradicts our assumption
about $\{\Znk \}$. We should also notice that if in (\ref{t}),
$t=\underline{+} \infty$ then in the proof of Lemma \ref{lem} for
$\la$ large enough we have $\{x=\frac{1}{2\la}\log \mid
\frac{b(\la)}{a(\la)} \mid\}\cap \text{int}(K')=\O$, and therefore
$Z|_{D_m^{\ep}}=0$. This means that we do not obtain any zero
lines in this canonical domain. In other words the zeros run away
from this canonical domain as $\la \to \infty$. But as we
mentioned in the introduction, all the Stokes lines on the real
axis are contained in the set of zero lines of every limit $Z$,
meaning that in Theorem \ref{FormOfZeros}, $\gamma$ is never
empty.

Now notice that because $\bigcup_{m}D_m^{\ep}$ covers the plane
except the $\ep$-neighborhoods around the turning points, we have
proved that
\\

$\qquad \qquad \qquad Z(\phi)=\frac{1}{\pi} \int_{\gamma} \phi(z)
|\sqrt{q(z)}|\,|d\gamma|, \qquad \phi \in C_{c}^{\infty}(\mathbb C
\backslash \bigcup_m B(z_m,\ep)).$
\\

To finish the proof we have to show that if $\phi_{\ep} \in
C_{c}^{\infty}(\bigcup_m B(z_m,\ep))$ is a bounded function of
$\ep$, then

$$ \lim_{\ep \to 0}\limsup_{\lank \to
\infty}\Znk(\phi_{\ep})=0.$$

This is clearly equivalent to showing that if $z_0$ is a turning
point, then

\begin{equation} \label{epsilon} \lim_{\ep \to 0}\limsup_{\la \to
\infty} \frac{\# \{z\in B(z_0,\ep)| \; y(z,\la)=0\}}{\la}=0.
\end{equation}

To prove this we use the following fact in \cite{F1} pages
$104-105$ or \cite{EF} pages $39-41$, which enables us to improve
the domain $D^{\ep}$ in (\ref{ep1}) and (\ref{ep2}) from a fixed
$\ep$ to $\ep(\la)$ dependent of $\la$ such that $\ep(\la)\to 0$
as $\la \to \infty$.

Let $D$ be a canonical domain with turning points $z_m$ on its
boundary. Assume $N(\la)$ is a positive function such that
$N(\infty)=\infty$. Now if we denote

$$ D(\la)= D \;\backslash \bigcup_{m} B(z_m,
|q'(z_m)|^{-1/3}N(\la)\la^{-2/3}),$$

Then in place of equations (\ref{ep1}) and (\ref{ep2}) we have

$$ \ep_1(z, \la), \; \ep_2(z,\la)=O( N(\la)^{-3/2}), \qquad
\text{uniformly in} \; D(\la), \quad \la \to \infty.$$

In fact this implies that Lemma \ref{lem} is true for every $\phi$
supported in $D$. This is because we can follow the proof of the
lemma line by line except that in (\ref{smallterm}) we get $\mu(z,
\la)=O(N(\la)^{-3/2}\la^{-1})$ uniformly in $D(\la)$ and
therefore, using $N(\infty)=\infty$, we can still conclude $
\text{mesh}(\wp) \to 0$ as $\la \to \infty$.

We choose $N(\la)=\la^{1/12}$. By the discussion in the last
paragraph in (\ref{epsilon}) we can replace $\ep$ by
$\ep(\la)=cN(\la)\la^{-2/3}=c\la^{-7/12}$, where
$c=|q'(z_0)|^{-1/3}$. Let us find a bound for the number of zeros
of $y(z, \la)$ in $B(z_0, \ep(\la))$. Let
$M=\text{sup}_{B(z_0,\delta)}(|q(z)|)$ where $\delta>0$ is fixed
and is chosen such that the ball $B(z_0,\delta)$ does not contain
any other turning points. We also choose $\la$ large enough so
that $\ep(\la)<\delta$. If $\zeta$ is a zero of $y(z, \la)$ in the
ball $B(z_0,\ep(\la))$ then by Corollary $11.1.1$ page $579$ of
\cite{H1} we know that there are no zeros of $y(z,\la)$ in the
ball of radius $\frac{\pi}{\sqrt{M}}\la^{-1}$ around $\zeta$
except $\zeta$. Therefore

$$ \#\{z \in B(z_0,\ep(\la))| \; y(z,\la)=0\} \leq
\frac{\text{area}(B(z_0,\ep(\la)+\frac{\pi}{2\sqrt{M}}\la^{-1}))}{\text{area}(B(\zeta,\frac{\pi}{2\sqrt{M}}\la^{-1}))}=O(
\la^{5/6}),$$

and so

$$\lim_{\la \to \infty}\frac{\#\{z \in B(z_0,\ep(\la))| \; y(z,\la)=0\}}{\la}=0. $$

This finishes the proof of Theorem \ref{FormOfZeros}.
\\

\begin{figure}
\begin{center} \includegraphics[scale=0.7]{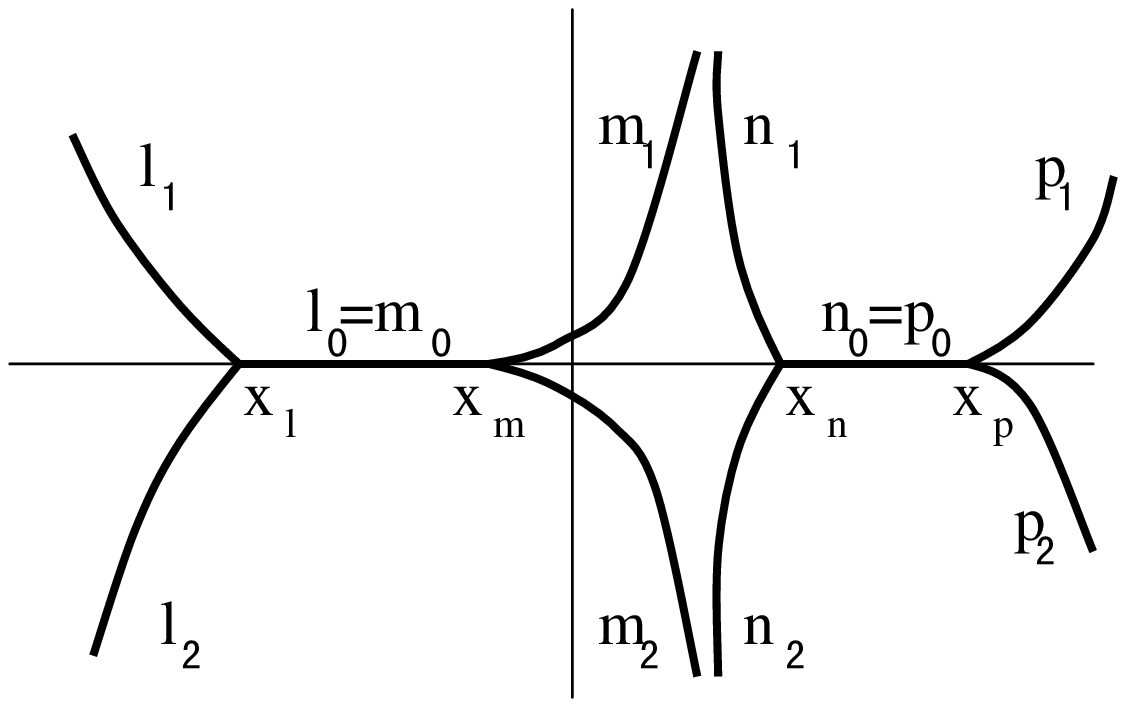}\end{center}
\caption{\label{DoubleWell}}
\end{figure}

\textsc{The Proofs of Theorems \ref{2well}, \ref{Nonsymm}}. We
will not prove Theorem \ref{1well}, because the proof is similar
to (in fact easier than) the proof of the two well potential. To
simplify our notations let us rename the turning points as
$x_l=a_0,x_m=a_1,x_n=a_2,x_p=a_3$. Then we can index the Stokes
lines as in Fig (\ref{DoubleWell}).

 We define the canonical domains $D_{l}$, $D_{m_0}$, $D$, $D_{n_0}$, and $D_{p}$  by

\begin{equation} \label{canonicaldomains} \begin{array}{ll}  l_1 \subset
D_l, \qquad m_1,m_0,l_2 \subset \partial D_l,
 \\

   m_0 \subset D_{m_0}, \qquad l_1,l_2,m_1,m_2 \subset
 \partial D_{m_0},
 \\

   m_1,n_1 \subset D, \qquad l_1,l_0,m_2,n_2,p_0,p_1 \subset \partial D,
 \\

   n_0 \subset D_{n_0}, \qquad p_1,p_2,n_1,n_2 \subset
 \partial D_{n_0},
 \\

  p_1 \subset D_p, \qquad n_1,n_0,p_2 \subset \partial D_p \; .
\end{array} \end{equation}

Notice that the complex conjugates of the these canonical domains
are also canonical domains and in fact if we include these complex
conjugates then we obtain a covering of the plane by canonical
domains. But because $q(x)$ is real, the zeros are symmetric with
respect to the $x$-axis, and it is therefore enough to find the
zeros in $D_l\cup D_{m_0}\cup D\cup D_{n_0}\cup D_p$. By lemma
\ref{lem} we only need to discuss the limit (\ref{t}) in each of
these canonical domains. First of all let us compute the equation
of the eigenvalues (\ref{bla}).

Here, the transition matrix $\Omega_{+,-}$, is the product of the
seven elementary
 matrices associated to the following sequence of triples:
 \\

$(D_p,l_1,x_p)\mapsto(D_{n_0},n_0,x_p)\mapsto(D_{n_0},n_0,x_n)\mapsto(D,n_1,x_n)$
$$\mapsto(D,m_1,x_m)
\mapsto(D_{m_0},m_0,x_m)\mapsto(D_{m_0},m_0,x_l)\mapsto(D_l,l_1,x_l).$$
In fact if we define

 $$ \al_1 =\int_{x_l}^{x_m} |\sqrt {q(t)}|dt, \quad \al_2= \int_{x_n}^{x_p} |\sqrt
 {q(t)}|dt, \quad \xi= \int_{x_m}^{x_n} \sqrt
 {q(t)}dt,$$ then by (\ref{Transition1}),(\ref{Transition2}) and (\ref{AlphaJ}), we have
\\

$\left(%
\begin{array}{c}
  a(\la) \\
  b(\la) \\
\end{array}%
\right) =\Omega_{+,-}\left(%
\begin{array}{c}
  0 \\
  1 \\
\end{array}%
\right)$

$$  =\left(%
\begin{array}{cc}
  0 & {\al_{l_0 l_1}^{-1}} \\
  1 & i\al _{l_1 1_2} \\
\end{array}%
\right)
\left(%
\begin{array}{cc}
  0 & e^{-i\la \al_1} \\
 e^{i\la \al_1}  & 0 \\
\end{array}%
\right)
\left(%
\begin{array}{cc}
  0 & \al _{m_1 m_0}^{-1} \\
  1 & i\al _{m_0 m_2} \\
\end{array}%
\right)
\left(%
\begin{array}{cc}
  e^{-\la \xi} & 0 \\
  0 & e^{\la \xi} \\
\end{array}%
\right)$$
$$ \qquad \qquad \qquad \left(%
 \begin{array}{cc}
  0 & \al _{n_0 n_1}^{-1} \\
  1 & i\al_{n_1 n_2}  \\
\end{array}%
\right)
\left(%
\begin{array}{cc}
  0 & e^{-i\la \al_2} \\
 e^{i\la \al_2} & 0 \\
\end{array}%
\right)
\left(%
\begin{array}{cc}
  0 & \al_{p_1 p_0}^{-1} \\
  1 & i\al_{p_0 p_2} \\
\end{array}%
\right) \left(%
\begin{array}{c}
  0 \\
  1 \\
\end{array}%
\right).$$

A simple calculation shows that
$$ b(\la)=\al_{p_1 p_0}^{-1} \al_{n_0 n_1}^{-1} e^{i\la (\al_2 -\al_1)}e^{-\la \xi}-
(\al_{m_0 m_2} e^{-i\la \al_1} + \al _{l_1 l_2} \al_{m_1 m_0}^{-1}
e^{i\la \al_1})(\al_{p_0 p_2} e^{-i\la \al_2}+ \al_{n_1 n_2}
\al_{p_1 p_0}^{-1} e^{i\la \al_2})e^{\la
  \xi}.$$

Hence $b(\la)=0$ implies that

\begin{equation} \label{eigenequation}\Gamma_1(\la)\Gamma_2(\la)= \al_{p_1 p_0}^{-1} \al_{n_0
n_1}^{-1} e^{i\la (\al_2 -\al_1)}e^{-2\la \xi}, \end{equation}

where \begin{equation}\label{Gamma1Gamma2} \begin{array}{ll}
\Gamma_1(\la)=\al_{m_0 m_2} e^{-i\la \al_1} + \al _{l_1 l_2}
\al_{m_1 m_0}^{-1} e^{i\la \al_1}=2\cos(\al_1
\la)+O(\frac{1}{\la}),
\\ \\

\Gamma_2(\la)= \al_{p_0 p_2} e^{-i\la \al_2}+ \al_{n_1 n_2}
\al_{p_1 p_0}^{-1} e^{i\la \al_2}=2\cos(\al_2
\la)+O(\frac{1}{\la}).
\\ \\

\end{array}
\end{equation}
Now let us discuss the limit in (\ref{t}) for each of the
canonical domains defined in (\ref{canonicaldomains}). Even though
the coefficients $a(\la)$, $b(\la)$ are different for different
canonical domains, we do not consider it in our notation.

By (\ref{y1princ}), (\ref{y2princ}) , (\ref{ep1}), and
(\ref{ep2}), it is clear that for $\la$ large enough there are no
zeros in $D_l^{\ep}$ and $D_p^{\ep}$. For $(D_{n_0},n_0,x_p)$ we
have
$$\left(%
\begin{array}{c}
  a(\la) \\
  b(\la) \\
\end{array}%
\right)=\left(%
\begin{array}{cc}
  0 & \al_{p_1 p_0}^{-1} \\
  1 & i\al_{p_0 p_2} \\
\end{array}%
\right)
\left(%
\begin{array}{c}
  0 \\
  1 \\
\end{array}%
\right)=\left(%
\begin{array}{c}
   \al_{p_1 p_0}^{-1} \\
  i\al_{p_0 p_2} \\
\end{array}%
\right).$$

Using (\ref{AlphaJ}),(\ref{Alpha12}), for the full sequence
$\la_n$ we have
$$\frac{1}{2\la_n} \log |\frac{b(\la_n)}{a(\la_n)}|=\frac{1}{2\la_n}
\log|\al_{p_1 p_2}|=0.$$

Hence $t=0$ and, by Lemma \ref{lem}, the Stokes line
$n_0=(a_2,a_3)$ is a zero line in $D_{n_0}$. The same proof shows
that the Stokes line $[a_0,a_1]$ is a zero line for the full
sequence $\Zn$ in $D_{m_0}$. Now it only remains to discuss the
limit in (\ref{t}) in the canonical domain $D$. For the triple
$(D,n_1,x_n)$ we have

$$\left(%
\begin{array}{c}
  a(\la) \\
  b(\la) \\
\end{array}%
\right)=\left(%
 \begin{array}{cc}
  0 & \al _{n_0 n_1}^{-1} \\
  1 & i\al_{n_1 n_2}  \\
\end{array}%
\right)
\left(%
\begin{array}{cc}
  0 & e^{-i\la \al_2} \\
 e^{i\la \al_2} & 0 \\
\end{array}%
\right)
\left(%
\begin{array}{cc}
  0 & \al_{p_1 p_0}^{-1} \\
  1 & i\al_{p_0 p_2} \\
\end{array}%
\right) \left(%
\begin{array}{c}
  0 \\
  1 \\
\end{array}%
\right)=\left(%
\begin{array}{c}
  e^{i\la \al_2}\al_{p_1 p_0}^{-1}\al _{n_0 n_1}^{-1} \\
  i\Gamma_2 (\la) \\
\end{array}%
\right).$$

Therefore, by the second equation in (\ref{AlphaJ}), we obtain

\begin{equation} \label{tD} t=\lim_{n \to \infty}\frac{1}{2\la_n}\log|\frac{i\Gamma_2 (\la_n)}{e^{i\la_n \al_2}\al_{p_1 p_0}^{-1}\al _{n_0
n_1}^{-1}}|=\lim_{n \to \infty}\frac{1}{2\la_n}\log|\Gamma_2
(\la_n)|.
\end{equation}

The limit (\ref{tD}) does not necessarily exist for the full
sequence $\{\la_n\}$. We study this limit in different cases as
follows:

\begin{enumerate}

\item {\boldmath $\frac{\al_1}{\al_2}=1$:}
\\

\noindent This is exactly the symmetric case in Theorem
\ref{2well}. It is easy to see that if $\al_1=\al_2$, then there
exists a translation on the real line which changes $q(z)$ to an
even function. When $q(z)$ is even, because of the symmetry in the
problem, we have $\Gamma_1(\la)=\Gamma_2(\la)$. On the other hand
equation (\ref{eigenequation}) implies that \begin{equation}
\label{eigenequation2} |\Gamma_1(\la)||\Gamma_2(\la)|=e^{-2\la
\xi}(1+O(\frac{1}{\la})). \end{equation}

\noindent This means that in the symmetric case, the full sequence
$\la_n$ satisfies
$$|\Gamma_1(\la_n)|=|\Gamma_2(\la_n)|=e^{-\la_n
\xi}(1+O(\frac{1}{\la_n})).$$

\noindent Therefore, by (\ref{tD}) we have $t=-\half \xi$ and
using the lemma the line $\Re S(a,z)=-\haf \xi$, is the zero line
in $D$. We note that this in fact determines the whole imaginary
axis, because $\Re S(a,0)=-\half \xi$. Also notice that in the
symmetric case, by our notations we have $a=a_2=x_n$. This proves
Theorem \ref{2well}.2.
\\

\item {\boldmath $\frac{\al_1}{\al_2}\neq1$:}
\\

\noindent In this case as we mentioned in the introduction, there
are more than one zero limit measures. Here the limit (\ref{tD})
behaves differently for the two subsequences in (\ref{Eigen1}) and
(\ref{Eigen2}) (notice that the equations (\ref{Eigen1}) and
(\ref{Eigen2}) in fact follow from (\ref{eigenequation})). It is
clear from (\ref{Gamma1Gamma2}) that if for a subsequence
$\{\lank\}$ we have a lower bound $\delta$ for $|\cos(\al_2
\lank)|$, then we have t=$\lim_{k \to \infty}
\frac{1}{2\lank}\log|\Gamma_2(\lank)|=0$. Also if we have a lower
bound $\delta$ for $|\cos(\al_1 \lank)|$ then $\lim_{k \to \infty}
\frac{1}{2\lank}\log|\Gamma_1(\lank)|=0$, and by
(\ref{eigenequation2}) we have $t=\lim_{k \to \infty}
\frac{1}{2\lank}\log|\Gamma_2(\lank)|=-\xi$. To find such
subsequences we denote for each $\ell=1,2$
$$ A_{\delta}^{(\ell)}= \{ \la_n ; \; |\cos(\al_{\ell}\la_n)| > \delta \}.$$
By (\ref{Eigen1}) and (\ref{Eigen2}), it is clear that up to some
finite sets $A_{\delta}^{(1)} \subset \{\la_n^{(2)}\}$ and
$A_{\delta}^{(2)} \subset \{\la_n^{(1)}\}$. We would like to find
the density of the subsets $A_{\delta}^{(1)}$ and
$A_{\delta}^{(2)}$ in $\{\la_n^{(2)}\}$ and $\{\la_n^{(1)}\}$
respectively. Here by the density of a subsequence $\{\lank\}$ of
$\{\la_n\}$ we mean

$$ d=\lim_{n \to \infty} \frac{\#\{k; \; \lank\leq \la_n\}}{n}. $$

\noindent If we set $\tau=\arcsin(\delta)$ then we have
$$A_{\delta}^{(1)}=\{n \in \mathbb N; \;
|(n+\half)\frac{\al_1}{\al_2}+(m+\half)|>\tau+O(\frac{1}{n}),\quad
\forall\; m \in \mathbb Z\},$$
$$A_{\delta}^{(2)}=\{n \in \mathbb N; \;
|(n+\half)\frac{\al_2}{\al_1}+(m+\half)|>\tau+O(\frac{1}{n}),\quad
\forall\; m\in \mathbb Z\}.$$

\noindent We only discuss the density of the subset
$A_{\delta}^{(1)}$. We rewrite this subset as

$$A_{\delta}^{(1)}=\{n \in \mathbb N; \;
|(2n+1)\al_1+(2m+1)\al_2|>2\al_2\tau+O(\frac{1}{n}),\quad
\forall\; m\in \mathbb Z\}.$$

\noindent From this we see that if $\frac{\al_1}{\al_2}$ is a
rational of the from $\frac{2r_1}{2r_2+1}$ (or
$\frac{2r_1+1}{2r_2}$), then because for every $m$ and $n$ we have
$$|(2n+1)(2r_1)+(2m+1)(2r_2+1))| \geq 1,$$ therefore $d(A_{\delta}^{(1)})=1$ for
$\tau=\frac{1}{8r_2+4}$. This proves Theorem \ref{Nonsymm}.2. When
$\frac{\al_1}{\al_2}$ is a rational of the form
$\frac{2r_1+1}{2r_2+1}$, we define
$$ B_{\delta}^{(1)}=\{n\in A_{\delta}^{(1)}; \; 2n+1\neq 0 \;
(\text{mod} \; 2r_2+1)\}.$$

\noindent Since for every $n\in B_{\delta}^{(1)}$ and $m\in
\mathbb Z$ we have
$$|(2n+1)(2r_1+1)+(2m+1)(2r_2+1))| \geq 1,$$
for $\tau=\frac{1}{8r_2+4}$, we get $d(A_{\delta}^{(1)})\geq
d(B_{\delta}^{(1)})=\frac{2r_2}{2r_2+1}.$ This completes the proof
of Theorem \ref{Nonsymm}.3.

\noindent To prove Theorem \ref{Nonsymm}.1, when
$\frac{\al_1}{\al_2}$ is irrational, we use the fact that the set
$\mathbb Z \al_1 \oplus \mathbb Z \al_2$ is dense in $\mathbb R$.
In fact it is easy to see that the subset $A=\{n\al_1+m\al_2|\;
n\in \mathbb N,\; m\in \mathbb Z\}$ is also dense. Now if we
rewrite $A_{\delta}^{(1)}$ as

$$ A_{\delta}^{(1)}=\{n \in \mathbb N; \;
|(n\al_1+m\al_2)+\half(\al_1+\al_2)|>\al_2\tau+O(\frac{1}{n}),\quad
\forall\; m \in \mathbb Z\},$$

\noindent
then from the denseness of the set $A$, it is not hard to
see that in this case
$d(A_{\delta}^{(1)})=1-\frac{2\al_2}{\al_1}\tau.$ Hence we
conclude that when $l=1$ there is a subsequence $\{\lankl\}$ of
$\{\lanl \}$ of density $1$. The same argument works for $l=2$.
This finishes the proof.
\end{enumerate}

\begin{figure}
\begin{overpic}[scale=0.6]{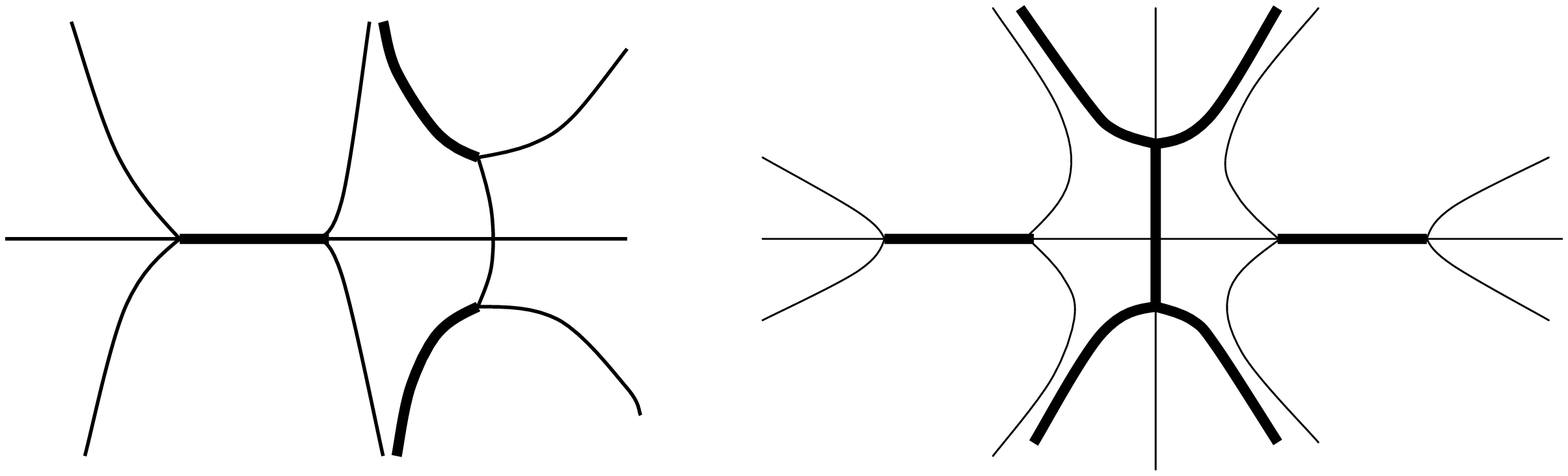}
\put(5,8){$q(z)=(z^2-a^2)(z^2+bz+c)$}
\put(52,8){$q(z)=(z^2-a^2)(z^2-b^2)(z^2+c^2)$}
\end{overpic}

\caption{\label{Degree6}}
\end{figure}

\begin{rem}\label{Remark}

In Figure (\ref{Degree6}) we have illustrated the zero lines for
the polynomials
$$q(z)=(z^2-a^2)(z^2+bz+c) \qquad \text{and}\qquad q(z)=(z^2-a^2)(z^2-b^2)(z^2+c^2).$$
The thickest lines in these figures are the zero lines. In fact
for these examples there is a unique zero limit measure as in the
other symmetric cases we mentioned in Theorems \ref{2well} and
\ref{1well}. We will not give the proofs, as they follow
similarly, but we would like to ask the following question:


Is there any polynomial potential with $n$ wells, $n\geq 3$ for
which there is a unique zero limit measure for the zeros of
eigenfunctions?

\end{rem}

\textbf{Acknowledgements:} I am sincerely grateful to Steve
Zelditch for introducing the problem and many helpful discussions
and suggestions on the subject.

\end{document}